\documentclass[aps,prl,preprint,showpacs,groupedaddress]{revtex4}
\usepackage{graphicx}
\usepackage{amsmath,bm,calc}
\usepackage[psamsfonts]{amssymb}

\begin{document}

% Use the \preprint command to place your local institutional report
% number in the upper righthand corner of the title page in preprint mode.
% Multiple \preprint commands are allowed.
% Use the 'preprintnumbers' class option to override journal defaults
% to display numbers if necessary
%\preprint{}

%Title of paper
\title{Effects of particle-number conservation
on heat capacity of nuclei}

% repeat the \author .. \affiliation  etc. as needed
% \email, \thanks, \homepage, \altaffiliation all apply to the current
% author. Explanatory text should go in the []'s, actual e-mail
% address or url should go in the {}'s for \email and \homepage.
% Please use the appropriate macro foreach each type of information

% \affiliation command applies to all authors since the last
% \affiliation command. The \affiliation command should follow the
% other information
% \affiliation can be followed by \email, \homepage, \thanks as well.
\author{K. Esashika}
\affiliation{Graduate School of Science and Technology, 
Chiba University, Inage, Chiba 263-8522, Japan}
\author{H. Nakada}
\email{nakada@faculty.chiba-u.jp}
%\homepage[]{Your web page}
%\thanks{}
%\altaffiliation{}
\affiliation{Department of Physics, Faculty of Science,
Chiba University, Inage, Chiba 263-8522, Japan}
\author{K. Tanabe}
\email{tanabe@phy.saitama-u.ac.jp}
\affiliation{Department of Physics, Faculty of Science, 
Saitama University, Sakura, Saitama 338-8570, Japan}

%Collaboration name if desired (requires use of superscriptaddress
%option in \documentclass). \noaffiliation is required (may also be
%used with the \author command).
%\collaboration can be followed by \email, \homepage, \thanks as well.
%\collaboration{}
%\noaffiliation

\date{\today}

\begin{abstract}
% insert abstract here
 By applying the particle-number projection
 to the finite-temperature BCS theory,
 the $S$-shaped heat capacity, which has recently been claimed
 to be a fingerprint of the superfluid-to-normal phase transition
 in nuclei, is reexamined.
 It is found that the particle-number (or number-parity) projection
 gives $S$-shapes in the heat capacity of nuclei
 which look qualitatively similar to the observed ones.
 These $S$-shapes are accounted for as effects
 of the particle-number conservation on the quasiparticle excitations,
% and occur even without the phase transition.
 and occur even when we keep the superfluidity at all temperatures
 by assuming a constant gap in the BCS theory.
% Although the observed $S$-shapes might still be correlated
% to the superfluid-to-normal phase transition,
% they cannot be exclusive evidence to the transition.
 The present study illustrates significance of the conservation laws
 in studying phase transitions of finite systems.
\end{abstract}

% insert suggested PACS numbers in braces on next line
\pacs{21.90.+f,21.60.-n,05.70.-a}
% insert suggested keywords - APS authors don't need to do this
%\keywords{heat capacity of nuclei, finite-temperature BCS theory,
% particle-number projection}

%\maketitle must follow title, authors, abstract, \pacs, and \keywords
\maketitle

% body of paper here - Use proper section commands
% References should be done using the \cite, \ref, and \label commands
%\section{}
% Put \label in argument of \section for cross-referencing
%\section{\label{}}
%\subsection{}
%\subsubsection{}

% If in two-column mode, this environment will change to single-column
% format so that long equations can be displayed. Use
% sparingly.
%\begin{widetext}
% put long equation here
%\end{widetext}

While phase transitions are striking phenomena in infinite systems,
it has been difficult to establish them in finite systems
such as atomic nuclei,
because their signatures are often obscured
by the quantum fluctuations~\cite{LA84}.
Since most nuclei have the superfluidity in their ground states,
\textit{i.e.} at zero temperature,
the superfluid-to-normal transition for increasing temperature
has been discussed theoretically~\cite{Dos95,RCR98,ERM00,LA01}.
Recently, high precision measurement
of nuclear level densities has been implemented~\cite{Sch01},
in which the level densities are extracted from the $\gamma$-ray data
with the help of the Brink-Axel hypothesis.
Converting the microcanonical information to the canonical one,
they have found that $S$-shapes appear in the graphs
of the heat capacity $C$ as a function of temperature $T$.
It has been argued that the $S$-shapes are a fingerprint
of the superfluid-to-normal phase transition,
since such $S$-shapes occur
if continuity is taken between heat capacity in the superfluid phase
(described by the constant-$T$ model)
and that in the normal fluid phase
(described by the backshifted Bethe formula).
Based on this picture, they estimated critical temperature
from their experimental data.
Similar $S$-shapes come out in theoretical calculations
including the quantum fluctuations~\cite{RCR98,LA01}.

The superfluid and the normal-fluid phases in nuclei
are defined within the mean-field picture;
in the Bardeen-Cooper-Schrieffer (BCS) theory
or in the Hartree-Fock-Bogoliubov theory.
In studying the superfluid-to-normal transition,
linkage to the mean-field picture should carefully be traced,
even when the $S$-shapes are reproduced by calculations
including a variety of quantum fluctuations.
Breakdown of a certain symmetry is often associated
with phase transitions.
In respect to the superfluid-to-normal transition,
the particle-number conservation is violated in the superfluid phase
while preserved in the normal-fluid phase,
within the mean-field theories.
However, this is an approximate picture
and the particle-number conservation is not violated
in actual nuclei,
restored via the quantum fluctuations.
Since restoration of the symmetry sometimes plays an important role
particularly in finite systems,
it is desired to investigate how the number conservation
affects the heat capacity of nuclei.

The so-called number-parity forms a subgroup isomorphic to $S_2$
of the $U(1)$ group accompanied by the particle-number.
The number-parity projection
in the finite-temperature BCS (FT-BCS) theory~\cite{Goo81}
was developed more than two decades ago~\cite{TSM81}.
The full number projection at finite temperature
is much more complicated task.
There is a fundamental difficulty
in the variation-after-projection scheme,
though an approximate solution
has recently been suggested~\cite{TN05}.
On the other hand, the number projection
in the variation-before-projection (VBP) scheme
was formulated in Refs.~\cite{TN05,EE93,RR94}.
In this paper, we apply the particle-number projection
as well as the number-parity projection in the FT-BCS theory,
and qualitatively investigate effects
of the particle-number conservation in the heat capacity of nuclei.
By the VBP scheme, we view effects only of the projected statistics,
without changing the excitation spectra given by the BCS Hamiltonian,
as a step of tracing effects of various quantum fluctuations.

We mainly consider the $^{161,162}$Dy nuclei.
Let us assume the following model Hamiltonian,
\begin{equation}
 \hat{H} = \hat{H}_p + \hat{H}_n\,,\quad
 \hat{H}_\tau = \sum_{k\in\tau} \varepsilon_k a_k^\dagger a_k
 - \frac{g_\tau}{4} \sum_{k,k'\in\tau} a_k^\dagger a_{\bar{k}}^\dagger
 a_{\bar{k'}} a_{k'}\quad(\tau=p,n)\,,
 \label{Hamil}
\end{equation}
where $\bar{k}$ indicates the time-reversal
of the single-particle (s.p.) state $k$.
The s.p. state $k$ and its energy $\varepsilon_k$ are determined
from the Nilsson model~\cite{BM2},
by assuming the quadrupole deformation
from the measured $E2$ strength~\cite{E2}.
We take $g_p=22/A$ and $g_n=27/A$\,MeV~\cite{KB}.
For the model space, we first define the Fermi energy
$\varepsilon_\mathrm{F}$ for each nucleus
by the arithmetic average
between the energy of the highest occupied Nilsson s.p. level
and that of the lowest unoccupied level,
without the residual interaction.
We then include all the s.p. levels satisfying
$|\varepsilon_k-\varepsilon_\mathrm{F}|<7\,\mathrm{MeV}$,
for both protons and neutrons.
The $g_\tau$ values are related to the model space;
with the present choice we reproduce the pairing gaps
of $\Delta_\tau\approx 12/\sqrt{A}\,\mathrm{MeV}$.
Although the Hamiltonian in Eq.~(\ref{Hamil}) is relatively simple,
it will be sufficient for qualitative study
of the pairing phase transition.
Note that the critical temperature of the deformed-to-spherical
shape phase transition is appreciably higher
than that of the superfluid-to-normal transition,
and that the s.p. levels hardly change
at $T\lesssim 1\,\mathrm{MeV}$~\cite{ERM00}.

In the FT-BCS theory, we introduce the auxiliary Hamiltonian
\begin{equation}
 \hat{H}' = \hat{H}'_p + \hat{H}'_n\,,\quad
 \hat{H}'_\tau = \hat{H}_\tau - \lambda_\tau \hat{N}_\tau
 \quad(\tau=p,n)\,,
\end{equation}
where $\lambda_\tau$ stands for the chemical potential
and $\hat{N}_\tau$ the number operator,
$\hat{N}_\tau = \sum_{k\in\tau} a_k^\dagger a_k$.
The quasiparticle (q.p.) operators are obtained
from the Bogoliubov transformation,
\begin{equation}
 \alpha_k = u_k a_k - v_k a_{\bar k}^\dagger\,, \label{Bog-tr}
\end{equation}
with $u_k^2+v_k^2=1$, and the q.p. energy is given by
\begin{equation}
 E_k = \sqrt{\tilde{\varepsilon}_k^2 + \Delta_\tau^2}\,;
 \quad\tilde{\varepsilon}_k = \varepsilon_k
 - {g_\tau} v_k^2 - \lambda_\tau\,,\quad
 \Delta_\tau = \frac{g_\tau}{2} \sum_{k\in\tau} u_k v_k\,. \label{qpe}
\end{equation}
The density operator in the FT-BCS theory is
\begin{equation}
 \hat{w}_0 = \frac{e^{-\hat{H}_0/T}}{\mathrm{Tr}(e^{-\hat{H}_0/T})}\,;
 \quad\hat{H}_0 = \sum_k E_k \alpha_k^\dagger \alpha_k\,.
 \label{stat0}
\end{equation}
Here $\mathrm{Tr}$ denotes the grand-canonical trace
in the model space.
The thermal expectation value of an observable $\hat{O}$
is calculated by $\langle\hat{O}\rangle_0
= \mathrm{Tr}(\hat{w}_0\hat{O})$.
Equation~(\ref{stat0}) is an approximation
of $\hat{H}'$ in the Boltzmann-Gibbs operator $e^{-\hat{H}'/T}$
by $(\mathrm{const.}+\hat{H}_0)$.
The entropy is defined by $S=-\mathrm{Tr}(\hat{w}_0\ln\hat{w}_0)$.
The FT-BCS equation,
which determines the $u_k$ and $v_k$ coefficients in Eq.~(\ref{Bog-tr}),
is obtained so as to minimize the grand potential
$\Omega=\langle\hat{H}'\rangle_0-TS$ for each $T$~\cite{Goo81}.
The chemical potential $\lambda_\tau$ is fixed
by the particle-number condition
$\langle\hat{N}_\tau\rangle_0=N_\tau$,
where $N_\tau$ is the particle number in the model space
corresponding to the specific nuclide.

Introducing the particle-number projector,
\begin{equation}
 \hat{P}_\tau = \frac{1}{2\pi} \int_{-\pi}^\pi d\varphi\,
  e^{-i\varphi(\hat{N}_\tau-N_\tau)}\,,
  \label{Nproj}
\end{equation}
we define the density operator in the projected statistics,
\begin{equation}
 \hat{w}_\mathrm{P} = \frac{\hat{P}_p\hat{P}_n e^{-\hat{H}_0/T}
  \hat{P}_p\hat{P}_n}{\mathrm{Tr}(\hat{P}_p\hat{P}_n e^{-\hat{H}_0/T})}\,.
 \label{stat-P}
\end{equation}
The thermal expectation value in the number-projected statistics
is obtained by $\langle\hat{O}\rangle_\mathrm{P}
= \mathrm{Tr}(\hat{w}_\mathrm{P}\hat{O})$.
The integration over $\varphi$ in Eq.~(\ref{Nproj})
takes account of a certain part of the two-body correlations
beyond the mean-field approximation.
In practical calculations,
the $\varphi$ integral is replaced by a discrete sum:
\begin{equation}
 \hat{P}_\tau =
  \frac{1}{M+1}\sum_{m=0}^M e^{-i\varphi_m(\hat{N}_\tau-N_\tau)}\,,
  \label{proj-sum}
\end{equation}
where $M$ stands for the number of the s.p. states
and $\varphi_m=2\pi m/(M+1)$.
If we set $M=1$ instead of the number of the s.p. states
in Eq.~(\ref{proj-sum}),
$\hat{P}_\tau$ is reduced to the number-parity projector~\cite{TN05},
\begin{equation}
 \hat{P}'_\tau = \frac{1}{2} \sum_{\varphi=0,\pi}
  e^{-i\varphi(\hat{N}_\tau-N_\tau)}\,.
\end{equation}
Correspondingly,
$\hat{w}_\mathrm{P}$ and $\langle\hat{O}\rangle_\mathrm{P}$
become those of the number-parity projection,
which we shall denote by $\hat{w}_{\mathrm{P}'}$
and $\langle\hat{O}\rangle_{\mathrm{P}'}
= \mathrm{Tr}(\hat{w}_{\mathrm{P}'}\hat{O})$.
In this paper we distinguish the unprojected,
the number-projected and the number-parity-projected
expectation values by the suffices as
$\langle\hat{O}\rangle_0$, $\langle\hat{O}\rangle_\mathrm{P}$
and $\langle\hat{O}\rangle_{\mathrm{P}'}$, respectively.

\begin{figure}
\includegraphics[scale=0.85]{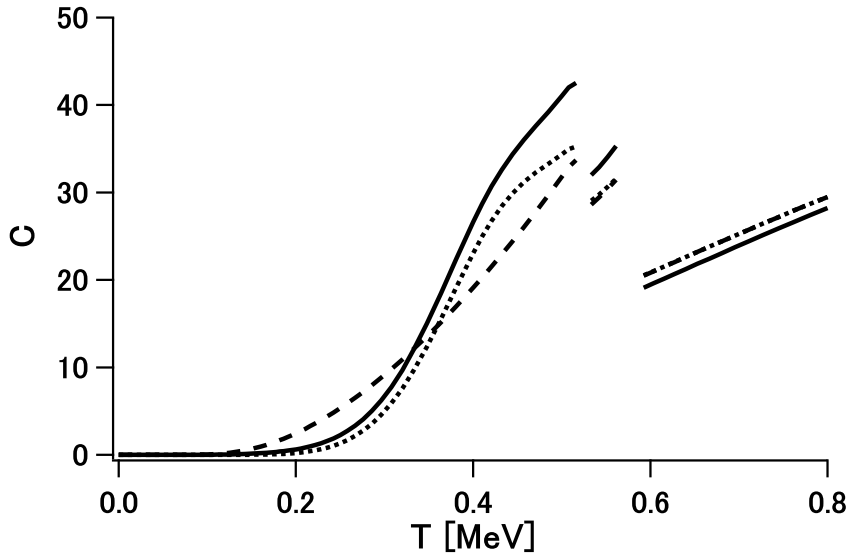}%
\caption{Heat capacity $C$ \textit{vs.} temperature $T$
 for $^{162}$Dy.
 The dashed line is obtained by the FT-BCS calculation without projection,
 while the solid and dotted lines are with the number
 and the number-parity projections, respectively.
 \label{Dy162t}}
\end{figure}

We here calculate heat capacity by $C=d\langle\hat{H}\rangle/dT$,
which is obtained by numerical differentiation
of $\langle\hat{H}\rangle$ for various $T$, in practice.
In Fig.~\ref{Dy162t},
the heat capacities with and without projection ($C_0$,
$C_{\mathrm{P}'}$ and $C_\mathrm{P}$) are depicted for $^{162}$Dy.
There occur two discontinuities in $C(T)$,
corresponding to the superfluid-to-normal transitions
for protons and neutrons.
This is confirmed by $\Delta_p$ and $\Delta_n$,
which rapidly vanish at the respective critical temperature $T_c$.
This signature to the transition does not disappear by the projection
because it is implemented after variation.
Although such discontinuities are unrealistic,
the present purpose is to investigate effects of the projection
mainly at $T<T_c$.
We have $T_c\approx 0.5-0.6\,\mathrm{MeV}$ in the present calculation,
slightly lower for protons than for neutrons.
In the normal fluid phase, the projections do not make
important differences;
difference of the number-parity-projected result
from the unprojected one is even almost invisible.
At $T<T_c$, we find that there is a certain effect
of the projection on $C(T)$,
either the number-parity or the number projection;
an $S$-shape appears when we apply the number or the number-parity
projection.
It is noted that, in the number-parity projection,
the zero-point of the energy may influence $C_{\mathrm{P}'}$,
since $\langle\hat{N}_\tau\rangle_{\mathrm{P}'}$ is displaced
from $N_\tau$ depending on $T$
because of the incomplete projection (after variation).
To suppress this influence,
we use $(\varepsilon_k-\varepsilon_\mathrm{F})$
instead of $\varepsilon_k$ in the Hamiltonian of Eq.~(\ref{Hamil}).

\begin{figure}
\includegraphics[scale=0.85]{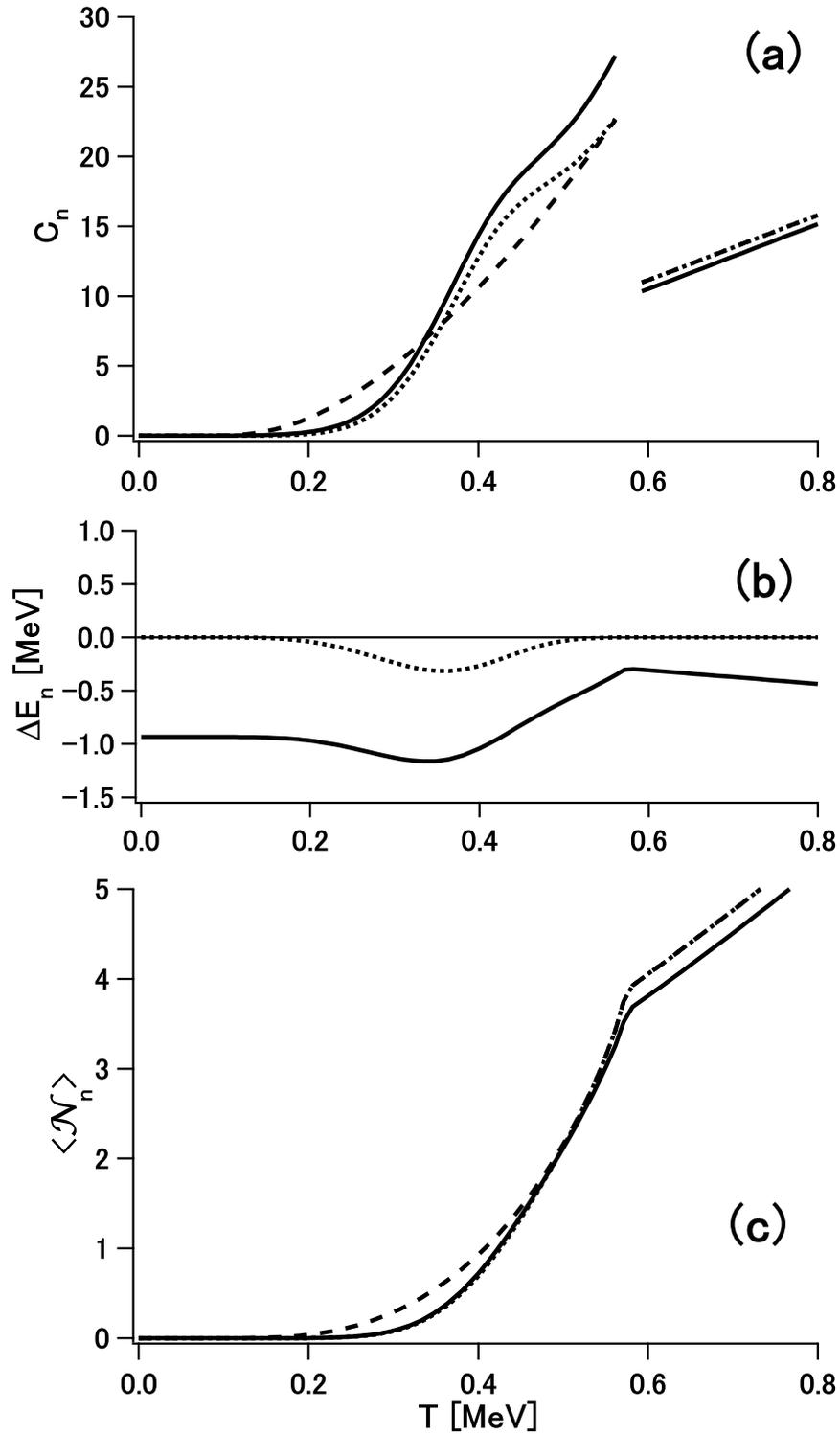}%
\caption{Thermal properties of the neutron part of $^{162}$Dy:
 (a) $C_n$, (b) $\mathit{\Delta}E_n$,
 (c) $\langle\hat{\mathcal{N}}_n\rangle$,
 as functions of the temperature.
 See Fig.~\ref{Dy162t} for conventions.
 \label{Dy162n}}
\end{figure}

Since the proton and neutron degrees-of-freedom
are separated in the Hamiltonian of Eq.~(\ref{Hamil}),
we have $C=C_p+C_n$, where $C_\tau=d\langle\hat{H}_\tau\rangle/dT$.
We present $C_n(T)$ for $^{162}$Dy in Fig.~\ref{Dy162n}(a),
so as to simplify discussions.
There is no qualitative difference between $C_p(T)$ and $C_n(T)$.
It is noticed that,
while $C_{n,0}=d\langle\hat{H}_n\rangle_0/dT$ increases
almost linearly at $(0.2\,\mathrm{MeV}\lesssim)T<T_c$,
an $S$-shape comes out both in $C_{n,\mathrm{P}'}=
d\langle\hat{H}_n\rangle_{\mathrm{P}'}/dT$
and $C_{n,\mathrm{P}}=d\langle\hat{H}_n\rangle_\mathrm{P}/dT$
in this temperature region.
In order to view effects of the projections more clearly,
we also show $\mathit{\Delta}E_{n,\mathrm{P}} =
\langle\hat{H}_n\rangle_\mathrm{P}-\langle\hat{H}_n\rangle_0$
and $\mathit{\Delta}E_{n,\mathrm{P}'} =
\langle\hat{H}_n\rangle_{\mathrm{P}'}-\langle\hat{H}_n\rangle_0$
as a function of $T$, in Fig.~\ref{Dy162n}(b).
Obviously, the slope in $\mathit{\Delta}E_{n,\mathrm{P}}(T)$ is equal
to the difference of $C_{n,\mathrm{P}}$ from $C_{n,0}$,
and likewise for $\mathit{\Delta}E_{n,\mathrm{P}'}(T)$.
In this respect, the $S$-shape in the number-parity-projected case
comes from the decrease of $\mathit{\Delta}E_{n,\mathrm{P}'}(T)$
at $0.2\lesssim T\lesssim 0.35\,\mathrm{MeV}$
as well as the increase at $0.35\lesssim T\lesssim 0.5\,\mathrm{MeV}$.
This behavior does not change by the full particle-number projection,
although $\mathit{\Delta}E_{n,\mathrm{P}}(T)$ shifts downward
to a certain extent, compared with $\mathit{\Delta}E_{n,\mathrm{P}'}(T)$.
We also plot in Fig.~\ref{Dy162n}(c)
the expectation value of the q.p. number
$\langle\hat{\mathcal{N}}_n\rangle$,
where $\hat{\mathcal{N}}_n=\sum_{k\in n}\alpha_k^\dagger\alpha_k$.
In Fig.~\ref{Dy162n}(c),
the number-parity-projected result
is almost indistinguishable from the fully number-projected result
at $T\lesssim 0.5\,\mathrm{MeV}$,
and from the unprojected one at $T\gtrsim 0.5\mathrm{MeV}$.
At $T\approx 0$, we have $\langle\hat{\mathcal{N}}_n\rangle\approx 0$
as it should be.
As $T$ goes up slightly to about $0.2\,\mathrm{MeV}$,
excitation to the 1~q.p. states gives rise to
increase of $\langle\hat{H}\rangle_0$,
and yields dominant contribution to $C_{n,0}$.
However, this is fictitious, resulting from the violation
of the number or the number-parity conservation.
Since the 1~q.p. states are removed,
$\langle\hat{H}\rangle_{\mathrm{P}'}$ tends to stay at its $T=0$ value,
giving the decrease of $\mathit{\Delta}E_{n,\mathrm{P}'}(T)$
and delaying the rise of $C_{n,\mathrm{P}'}(T)$.
As $T$ grows further, excitation to the 2~q.p. states starts
contributing to $\langle\hat{H}\rangle_0$,
and also to $\langle\hat{H}\rangle_{\mathrm{P}'}$.
If more and more q.p.'s are excited,
the q.p. number distributes broadly, and
the lack of the odd number-parity states is no longer important.
Then $\mathit{\Delta}E_{n,\mathrm{P}'}$ becomes vanishing again,
and $C_{n,\mathrm{P}'}$ also approaches $C_{n,0}$.
This feature is inherited in the number-projected result.
Thus the occurrence of the $S$-shape in $C(T)$ is accounted for
as effects of the number (or number-parity) conservation
on the q.p. excitations.

\begin{figure}
\includegraphics[scale=0.85]{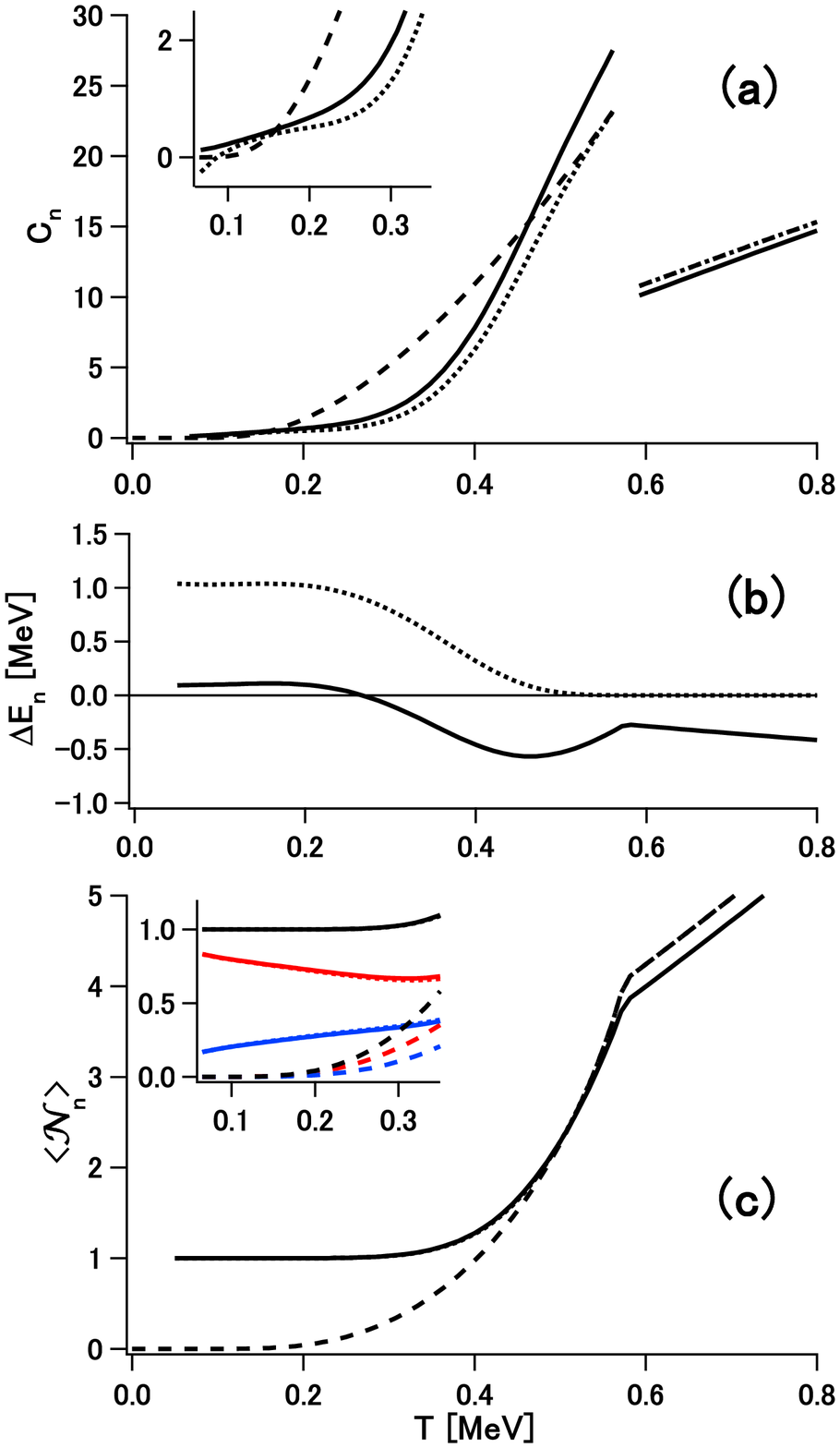}%
\caption{Thermal properties of the neutron part of $^{161}$Dy:
 (a) $C_n$, (b) $\mathit{\Delta}E_n$,
 (c) $\langle\hat{\mathcal{N}}_n\rangle$.
 $C_n$ at $0.07<T<0.35\,\mathrm{MeV}$ is
 amplified in the inset to (a).
 In the inset to (c),
 $\langle\sum_{k\in\mathrm{SP1}_n}
 \alpha_k^\dagger\alpha_k\rangle$ (red lines)
 and $\langle\sum_{k\in\mathrm{SP2}_n}
 \alpha_k^\dagger\alpha_k\rangle$ (blue lines)
 are also presented as well as the total q.p. number (black lines),
 where $\mathrm{SP1}_n$ is composed of
 the Nilsson s.p. level with $\varepsilon_\mathrm{F}$ and
 the neighboring two (one higher and one lower) levels,
 while $\mathrm{SP2}_n$ consists of
 the next neighboring six (three higher and three lower) levels.
 See Fig.~\ref{Dy162t} for other conventions.
 \label{Dy161n}}
\end{figure}

In $^{161}$Dy, the proton degrees-of-freedom are almost the same
as in $^{162}$Dy.
$C_n(T)$, $\mathit{\Delta}E_n(T)$ and
$\langle\hat{\mathcal{N}_n}\rangle$ in $^{161}$Dy
are presented in Fig.~\ref{Dy161n}.
In the graph of $C_n(T)$, we find an $S$-shape somewhat similar to
that in $^{162}$Dy, as a result of the projection.
However, the $S$-shape in the projected result is less conspicuous,
particularly around $T=T_c(\approx 0.6\,\mathrm{MeV})$.
It is remarked that similar even-odd difference is observed
in the experiments~\cite{Sch01}
and in more realistic calculations~\cite{RCR98,LA01}.
Moreover, we view small but non-vanishing values at low $T$
($0.05\lesssim T\lesssim 0.2\,\mathrm{MeV}$)
in the projected results (see inset to Fig.~\ref{Dy161n}(a)).
These even-odd differences are again accounted for
in terms of the q.p. excitation picture.
Since the neutron number is odd in $^{161}$Dy,
the q.p. number should be 1 at $T\approx 0$,
which is not taken into account in the unprojected result.
Therefore, $\mathit{\Delta}E_n$ is higher
by about $\Delta_n(\approx 1\,\mathrm{MeV})$
at $T\approx 0$ than in $^{162}$Dy.
As $T$ goes up, $\mathit{\Delta}E_{n,\mathrm{P}'}$ decreases,
because excitation to the 2~q.p. states predominantly contributes
to $\langle\hat{H}\rangle_0$,
which is eliminated in $\langle\hat{H}\rangle_{\mathrm{P}'}$.
For $T\gtrsim 0.5\,\mathrm{MeV}$, many q.p. states mix up
and the lack of even number q.p. states becomes less important.
This leads to $\mathit{\Delta}E_{n,\mathrm{P}'}\approx 0$.
However, since $\mathit{\Delta}E_{n,\mathrm{P}'}(T=0)$ is higher
than in $^{162}$Dy,
$\mathit{\Delta}E_{n,\mathrm{P}'}\approx 0$
at $T\gtrsim 0.5\,\mathrm{MeV}$ implies that
$\mathit{\Delta}E_{n,\mathrm{P}'}$ does not go up to a great extent
at $T\approx T_c$,
making the upper bend in the $S$-shape of $C_{n,\mathrm{P}'}(T)$ weak.
Although $\mathit{\Delta}E_{n,\mathrm{P}}$ shifts downward,
the full number projection gives analogous behavior.
The structure in $C_n(T)$ at low $T$ is also connected
to the q.p. excitation.
While there is only a single 0~q.p. state present in the even systems,
there are several 1~q.p. states having close energy to one another.
Hence, excitation from the lowest-lying 1~q.p. state
to the higher-lying 1~q.p. states is possible even at low $T$.
Note that this effect is only the cases for the projected statistics.
To illustrate this mechanism, the q.p. number
corresponding to the three Nilsson s.p. levels
adjacent to $\varepsilon_\mathrm{F}$
and that for the next nearest six levels
(three higher and three lower levels)
are separately depicted in the inset to Fig.~\ref{Dy161n}(c).
Although $\langle\hat{\mathcal{N}}_n\rangle_\mathrm{P}$ or
$\langle\hat{\mathcal{N}}_n\rangle_{\mathrm{P}'}$ stays unity
at $T\lesssim 0.3\,\mathrm{MeV}$,
we view that excitation among the s.p. states occurs,
giving small but non-negligible heat capacity.

As expected from the q.p. excitation picture,
the above results are insensitive to nuclide,
except that $C(T)$ at low $T(\lesssim 0.3\,\mathrm{MeV})$
for odd nuclei somewhat depends on the s.p. levels
around $\varepsilon_\mathrm{F}$.
This has been confirmed
by calculations for neighboring Dy isotopes~\cite{M-thesis}.
The q.p. excitation mechanism is so generic
that it should not depend on the nuclear shapes,
which has also been confirmed by calculations
for Sn isotopes~\cite{M-thesis}.

\begin{figure}
\includegraphics[scale=0.85]{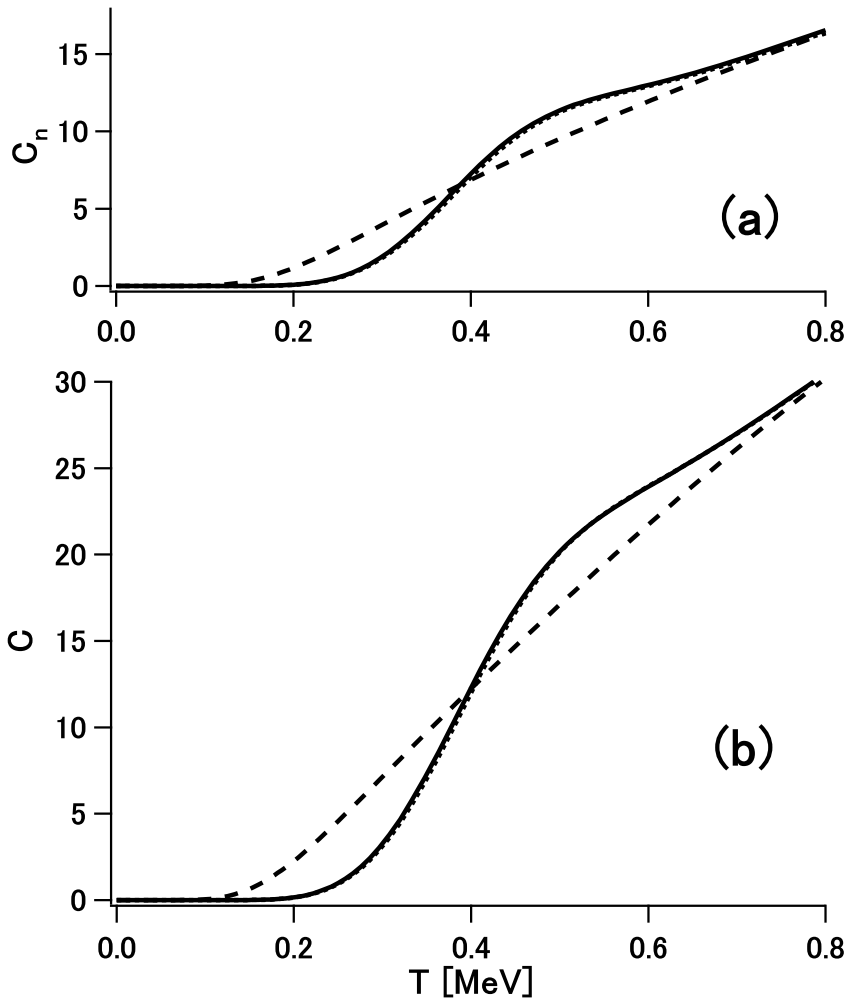}%
\caption{Heat capacities for $^{162}$Dy: (a) $C_n$ and (b) $C$,
 with $\tilde{\varepsilon}_k$ and $\Delta_\tau$
 fixed to be the $T=0$ values.
 \label{all-super}}
\end{figure}

\begin{figure}
\includegraphics[scale=0.85]{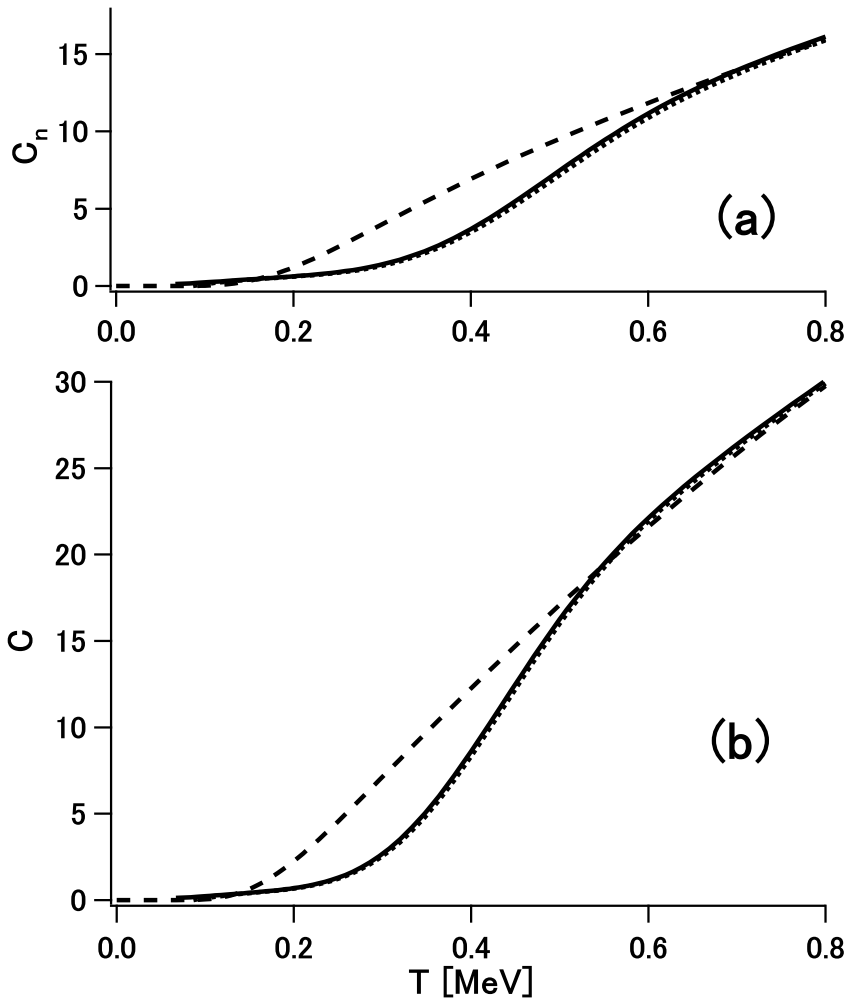}%
\caption{Heat capacities for $^{161}$Dy: (a) $C_n$ and (b) $C$,
 with $\tilde{\varepsilon}_k$ and $\Delta_\tau$
 fixed to be the $T=0$ values.
 \label{all-super2}}
\end{figure}

The number (or number-parity) conservation thus gives rise to
the $S$-shapes in the heat capacity and their even-odd differences.
Since it is explained within the q.p. excitation picture,
the $S$-shaped heat capacity does not seem straightforwardly linked
to the superfluid-to-normal transition.
To investigate this point further,
we try the following calculation:
instead of solving the FT-BCS equation at each $T$,
we keep using the solution at $T=0$
for $\tilde{\varepsilon}_k$ and $\Delta_\tau$,
and therefore for $u_k$ and $v_k$.
By this treatment the nucleus stays in the superfluid phase
at any $T$.
The $T$-dependence enters only via the explicit one
in the Boltzmann-Gibbs operator.
Resultant $C_n$ and $C=C_p+C_n$
are shown in Fig.~\ref{all-super} for $^{162}$Dy,
and in Fig.~\ref{all-super2} for $^{161}$Dy.
%We have the $S$-shapes even without the phase transition.
We have the $S$-shapes even though the superfluidity is kept
at any $T$.
The even-odd differences in $C$ are also viewed
in this artificial model.


The observed $S$-shapes in $C(T)$ are deduced from
the energy-dependence of the level densities~\cite{Sch01}.
We have confirmed that the present calculations with projection
are qualitatively consistent with the measured level densities,
whether or not $\hat{H}_0$ is $T$-dependent.
However, the present results indicate
a mechanism of the $S$-shapes in $C(T)$
different from the picture assumed in Ref.~\cite{Sch01},
in which the $S$-shapes emerge via continuity between the two phases.
Still, one should not immediately conclude
that the $S$-shapes shown in the present work
are irrelevant to the transition.
The $T_c$ value in the present FT-BCS calculations is close
to the temperature region where the $S$-shapes appear,
as in the experiments,
which may suggest that the $S$-shapes might have correlation
to the transition,
even though their connection is not straightforward.
For instance, suppose that the mixing of many q.p. states,
which causes the upper bend of the $S$-shape in $C(T)$,
also drives the phase transition in the BCS approximation,
the $S$-shapes are indirectly correlated to the transition.
It has been shown that the $S$-shapes take place
(and the discontinuity at critical temperature is washed out)
in realistic calculations~\cite{RCR98,LA01},
in which various correlations are taken into account.
However, in practice, relation of the $S$-shapes in $C(T)$
to the superfluid-to-normal phase transition
has not yet been clear enough.
It is desired to inspect carefully
how individual quantum fluctuations affect the heat capacities of nuclei.
We just point out here that the number conservation should play
a certain role in producing the $S$-shapes,
even though it may not be a full explanation.
It is emphasized that effects of the conservation laws
should not be discarded
in discussing phase transitions in finite systems.

In summary,
we have reexamined the heat-capacity of nuclei,
applying the particle-number projection
to the finite-temperature BCS theory.
Since the projection is carried out after the variation,
the sharp discontinuity at critical temperature remains
which is not observed in experiments.
While the $S$-shapes in the heat capacity have been claimed
to be a fingerprint of the superfluid-to-normal phase transition,
it is found that the particle-number projection
gives $S$-shapes in the heat capacity of nuclei
analogous to the observed ones,
apart from the discontinuity.
The even-odd difference in the heat capacity
is also produced by the projection.
Except low $T$ part of odd nuclei,
the number-parity projection gives similar heat capacity
to the full number projection,
if the s.p. energies are appropriately shifted.
These $S$-shapes are accounted for
in terms of the quasiparticle excitations,
%and occur even without the transition.
and occur even when we keep the superfluidity at all temperatures
by assuming a constant gap in the BCS theory.
Although the observed $S$-shapes could still correlate
to the phase transition,
their relation should be inspected carefully.
The even-odd difference is also understood
in the context of the quasiparticle excitations,
in which the particle-number (or the number-parity) conservation
plays a crucial role.
The present study reveals
significant role of the particle-number conservation
in the heat capacity of nuclei.

\begin{acknowledgments}
% put your acknowledgments here.
 The present work is financially supported
 as Grant-in-Aid for Scientific Research (B), No.~15340070,
 by the Ministry of Education, Culture, Sports, Science and Technology,
 Japan.
 Numerical calculations are performed on HITAC SR8000
 at Information Initiative Center, Hokkaido University.
\end{acknowledgments}

% Create the reference section using BibTeX:
%\bibliography{basename of .bib file}

\begin{thebibliography}{99}
 \bibitem{LA84} S. Levit and Y. Alhassid,
	 Nucl. Phys. \textbf{A413}, 439 (1984);
	 H. Nakada and Y. Alhassid,
	 Phys. Rev. Lett. \textbf{79}, 2939 (1997).
 \bibitem{Dos95} T. D$\o$ssing \textit{et al.},
	 Phys. Rev. Lett. \textbf{75}, 1276 (1995).
 \bibitem{RCR98} R. Rossignoli, N. Canosa and P. Ring,
	 Phys. Rev. Lett. \textbf{80}, 1853 (1998).
 \bibitem{ERM00} J. L. Egido, L. M. Robledo and V. Martin,
	 Phys. Rev. Lett. \textbf{85}, 26 (2000).
 \bibitem{LA01} S. Liu and Y. Alhassid,
	 Phys. Rev. Lett. \textbf{87}, 022501 (2001).
 \bibitem{Sch01} A. Schiller \textit{et al.},
	 Phys. Rev. C \textbf{63}, 021306(R) (2001);
	 M. Guttormsen \textit{et al.},
	 Phys. Rev. C \textbf{68}, 064306 (2003).
 \bibitem{Goo81} A. L. Goodman, Nucl. Phys. \textbf{A352}, 30 (1981).
 \bibitem{TSM81} K. Tanabe, K. Sugawara-Tanabe and H. J. Mang,
	 Nucl. Phys. \textbf{A357}, 20 (1981).
 \bibitem{TN05} K. Tanabe and H. Nakada,
	 Phys. Rev. C \textbf{71}, 024314 (2005).
 \bibitem{EE93} C. Esebbag and J. L. Egido,
	 Nucl. Phys. \textbf{A552}, 205 (1993).
 \bibitem{RR94} R. Rossignoli and P. Ring,
	 Ann. Phys. (N.Y.) \textbf{235}, 350 (1994).
 \bibitem{BM2} A. Bohr and B. R. Mottelson,
	 \textit{Nuclear Structure} vol.~2 (Benjamin, New York, 1975),
	 Chap.~5. 
 \bibitem{E2} S. Raman, C. W. Nestor Jr. and P. Tikkanen,
	 At. Data Nucl. Data Tables \textbf{78}, 1 (2001).
 \bibitem{KB} M. Baranger and K. Kumar,
	 Nucl. Phys. \textbf{A110}, 490 (1968).
 \bibitem{M-thesis} K. Esashika, Master thesis, Chiba University, 2005
	 (unpublished, in Japanese).
\end{thebibliography}

\end{document}